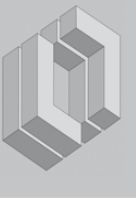

# ON THE SIMULATION OF THE SEISMIC ENERGY TRANSMISSION MECHANISMS

*Giulio Zuccaro*[1,2], *Daniela De Gregorio*[2], *Magdalini Titirla*[3],
*Mariano Modano*[1], *Luciano Rosati*[1]

[1] Department of Structures for Engineering and Architecture,
University of Naples Federico II, Napoli, Italy

[2] LUPT-PLINIVS Study Centre, University of Naples Federico II, Napoli, Italy

[3] Laboratory of Composite Materials for Construction LMC2, University Claude Bernard Lyon 1, Lyon, France

SUMMARY: *In recent years, considerable attention has been paid to research and development methods able to assess the seismic energy propagation on the territory. The seismic energy propagation is strongly related to the complexity of the source and it is affected by the attenuation and the scattering effects along the path. Thus, the effect of the earthquake is the result of a complex interaction between the signal emitted by the source and the propagation effects.*

*The purpose of this work is to develop a methodology able to reproduce the propagation law of seismic energy, hypothesizing the "transmission" mechanisms that preside over the distribution of seismic effects on the territory, by means of a structural optimization process with a predetermined energy distribution.*

*Briefly, the approach, based on a deterministic physical model, determines an objective correction of the detected distributions of seismic intensity on the soil, forcing the compatibility of the observed data with the physical-mechanical model. It is based on two hypotheses: (1) the earthquake at the epicentre is simulated by means of a system of distortions split into three parameters; (2) the intensity is considered coincident to the density of elastic energy. The optimal distribution of the beams stiffness is achieved, by reducing the difference between the values of intensity distribution computed on the mesh and those observed during four regional events historically reported concerning the Campania region (Italy).*

## 1    Introduction

After several seismic events, anomalous distributions of buildings damage, with strong differences between adjacent areas, can be observed. It can be caused by different vulnerability of buildings, as well as by changes in the characteristics of ground motion due to local geological and geomorphological conditions.

It is well known that the seismic wave propagation is influenced by: the characteristics of the earthquake excitation (magnitude, type of fault breaking and source-site distance) and, the

---

Corresponding author: Giulio Zuccaro, Department of Structures for Engineering and Architecture,
University of Naples Federico II, Napoli, Italy.
Email: zuccaro@unina.it

local response in function of very complex interactions between seismic waves and local conditions, generally known as "site effects". More specifically, spatial variation in seismic ground motions is manifested as measurable differences in amplitude and phase of seismic motions recorded over extended areas. It has an important effect on the response of lifelines, such as bridges, because these structures extend over long distances parallel to the ground and their supports undergo differential motions during an earthquake (Fontara et al, 2015; Fontara et al, 2017).

A challenge for the research networks of different areas has been placed worldwide with regard to evaluate the seismic attenuation laws, in order to determine the seismic motion in a given area, once the motion in a reference site is known. The formulation of physical-mathematical models that represent the actual complexity of the phenomenon and their resolution, has considerable difficulties. With the exception of a very limited number of simple situations, the physical-mathematical approach does not lead to solutions in a closed form, that is, integrally analytical (Douglas, 2014 and 2017). Therefore, in general it is necessary to adopt numerical procedures, often with approaches based on continuous discretization processes.

Various researchers have published reviews of ground-motion simulation techniques (e.g. Aki, 1982; Shinozuka, 1988; Anderson, 1991; Erdik and Durukal, 2003; Douglas and Aochi, 2008). The models for the prediction of earthquake ground motions are based on two approaches (Ólafsson et al., 2001): (1) the 'mathematical' ones, where the model is analytically based on physical principles; and (2) the 'experimental' ones, where the mathematical model, which is not necessarily based on physical insight, is fitted to experimental data. In addition, there are hybrid approaches combining elements of both approaches.

Among bidimensional (2D) numerical procedures, the Finite Element Method (FEM) is frequently used to assess the seismic wave propagation (Lysmer and Drake, 1972; Bao et al., 1998; Ma et al., 2007). The continuous domain (or reference site") is dividing into an equivalent system of smaller subdomains (of volume, surface and beam),.taking into account the soil heterogeneity, the boundary conditions and the not linear behaviour of materials.

In this perspective, the present research activity is developed. The main objective is to furnish a methodology able to define the propagation modalities of the seismic energy, through a structural optimization procedure with a predetermined energy distribution (Baratta and Zuccaro, 1990). The approach, based on a physical model of a deterministic type, determines the seismic intensity distribution on the territory, based on objective correction, forcing the compatibility of the observed data with the physical-mechanical model.

## 2   Problem definition

The seismic event is identified with the elastic energy $E(e,s)$, symmetric, accumulated in proximity of the site "$s$" caused by a seismic event at the epicentre "$e$"; the measure of the seismic event is determined by suitably correlating said energy with an intensity grading, according to any macro-seismic scale (MCS, MSK, EMS). The investigated soil, assimilated to continuous elastic, is discretized into a plane reticular structure formed by braced square meshes with side of approximately 10Km (Figure 1).

The solution to the problem is pursued by researching the function $E(e,s)$, applying the laws of reciprocity which govern the theory of elasticity, with the object of obtaining a prefixed distribution of energy, coincident with the aforesaid, deriving from macro-seismic intensities observed during previous seismic events. In practice, we must resolve a problem of structural optimizing at known geometry. In the present study, the Campania-Lucania territory is investigated.

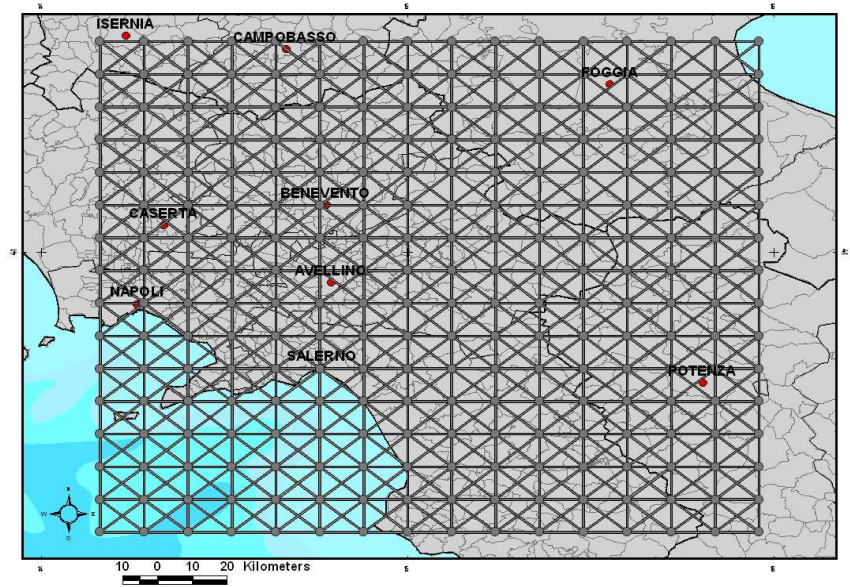

Figure 1 - *Discretization of the area of interest, equivalent truss network made of by 930 strokes connected by 256 nodes.*

From the Isoseistic Atlas of the P.F.G.-C.N.R. (1985) four seismic events registered during the 20th century in the Campania-Lucania Appennine region have been located (Figure 2). For each seismic event, from the quoted plane of the intensities, the values of the nodal intensities $I_{oi}$ of the equivalent reticular structure have been calculated, averaging the distance of the node *i-th* from the closer isoseismic curves, that is, by interpolating linearly. As may be noted, the calculated values of the observed nodal intensities $I_{oi}$ are not expressed by integer numbers - as defined by any macro-seismic intensity scale (MCS, MSK, and EMS). However such an assumption turns out to be less arbitrary if we consider the intensity as a measure of the propagated seismic energy. Such observed intensities $I_{oi}$ are correlated to the elastic energy $\phi_{oi}$ stored in proximity of the generic node of the mesh, through the empiric relation:

$$\phi_{oi} = 10^{(I_{oi}-5)} \tag{1}$$

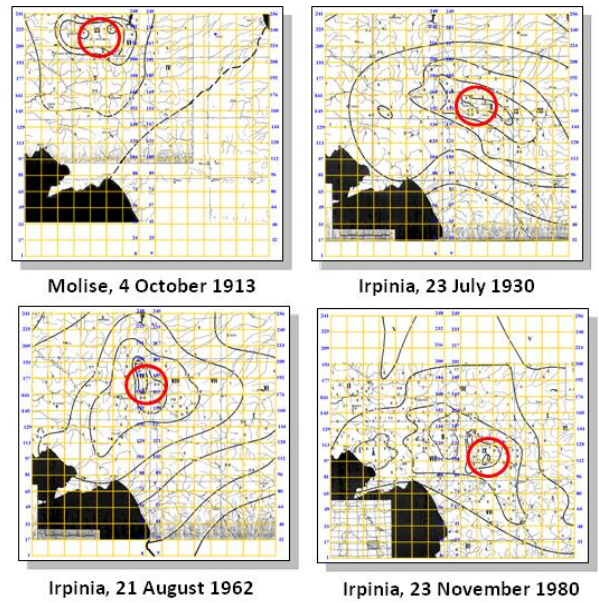

Figure 2 - *Isoseismic maps of the related events and numbering of the mesh nodes (the epicentral cell is circled in red).*

## 3    Analytical treatment

For each seismic event, we first locate within the mesh the Cell nearest to the epicentral area, and we simulate the seismic event by a self-balanced system of distorting parameters, applied to the 4 nodes defining the cell (**Figure 3**).

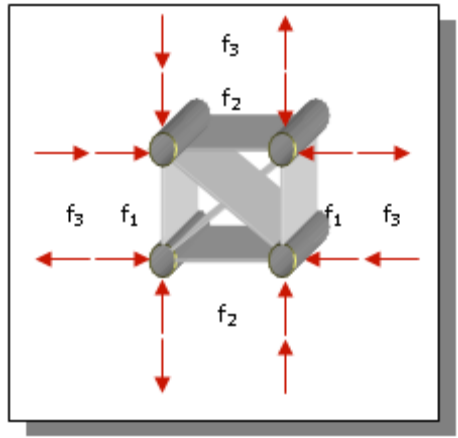

Figure 3 - *System of distorting parameters acting on the epicentral cell.*

Detailed description of the terms that define the problem is given in Table 1.

Table 1 - *Variants assumed in the problem solution.*

| Sign | Description |
|---|---|
| $\bar{n} = 256$ | Number of nodes in the mesh |
| N = 930 | Number of beams |
| $n = (2\bar{n} - 3)$ | Degrees of freedom |
| $\mathbf{B}_{(N,n)}$ | Compatibility matrix |
| $\boldsymbol{\omega}_{k(n)} \quad k = 1,2,3$ | Unit vector of the distortive epicentral parameters |
| $\varphi_{(\bar{n})}$ | Vector of the nodal equivalent energies |
| $\mathbf{u}_{(n)}$ | Vector of the free nodal displacements |
| $\boldsymbol{\Delta l}_{(N)} = \mathbf{B}_{(N,n)} \mathbf{u}_{(n)}$ | Vector of the beam extensions |
| $\mathbf{D}_{(N,N)}$ | (Constitutive low Matrix) Material Property Matrix |
| $\mathbf{K}_{(N,N)}$ | Diagonal stiffness matrix |
| $\mathbf{n}_{(N)} = \mathbf{D}_{(N,N)} \boldsymbol{\Delta l}_{(N)}$ | Vector of the axial strains of the beams |
| $\mathbf{y}_{(N-n)}$ | Vector of the hyperstatic axial strains |
| $\mathbf{t}_{(N-\bar{n})}$ | Vector of the free variables of energetic collimation |
| $\mathbf{x}_{(N)}$ | Vector of the energies stored in the beams |
| $f_k \quad con \quad k = 1,2,3$ | Distortive parameters equivalent to the seismic event |
| $\mathbf{f}_{(n)} = \sum_{k=1}^{3} f_k \boldsymbol{\omega}_{k(n)}$ | Vector of the nodal strengths |

Denoting by *N* the number of beams, we know the variables $\mathbf{B}_{(N,n)}$, $\boldsymbol{\omega}_{k(n)}$, $\varphi_{(\bar{n})}$ $\boldsymbol{\Delta l}_{(N)}, n_{(N)}$, $\mathbf{D}_{(N,N)}, \mathbf{u}_{(n)}, \mathbf{y}_{(N-n)}, \mathbf{t}_{(N-\bar{n})}, \mathbf{x}_{(N)}, \mathbf{f}_{(3)}$ are unknown. At solution, the previous variables have to fulfil equilibrium and compatibility of the deformed configuration, compatibly with a suitable simulation of the assigned energy distribution.

## 4 Equilibrium and compatibility

Equilibrium is enforced by setting:

$$\mathbf{B}^T_{(n,N)} \mathbf{n}_{(N)} = \mathbf{f}_{(n)} \tag{2}$$

Which, as already defined in the table 1 is equals to:

$$\mathbf{B}^T_{(n,N)} \mathbf{n}_{(N)} = \sum_{k=1}^{3} f_k \boldsymbol{\omega}_{k(n)} \tag{3}$$

Eq. (2) represents a system of "*n*" equations in the "*N*" unknown **n**. However, in general, the number of the unknowns is greater than the number of the equilibrium equations, (*N>n*) so that the problem is hyperstatic; in order to determine the hyperstatic unknowns. A partition of the compatibility matrix can be done as follows:

$$\mathbf{B}^T_{(n,N)} = \left[ \mathbf{B}^T_{1(n,n)} \quad \vdots \quad \mathbf{B}^T_{2(n,N-n)} \right] \tag{4}$$

in which the first "$n \times n$" lines the matrix $\mathbf{B}^T_{1(n,n)}$ is square and non-singular, while the second one $\mathbf{B}^T_{2(n,N-n)}$ is rectangular.

The same operation can be performed on the axial strain vector **n** collecting the first "$n$" entries in a vector $\mathbf{n}_1$ and the additional "$N$-$n$" entries in a vector $\mathbf{n}_2$:

$$\mathbf{n}_{(N)} = \begin{bmatrix} \mathbf{n}_{1(n)} \\ \mathbf{n}_{2(N-n)} \end{bmatrix} \tag{5}$$

Thus eq. (2) is partitioned as follows:

$$\left[ \mathbf{B}^T_{1(n,n)} \quad \mathbf{B}^T_{2(n,N-n)} \right] \begin{bmatrix} \mathbf{n}_{1(n)} \\ \mathbf{n}_{2(N-n)} \end{bmatrix} = \sum_{k=1}^{3} f_k \boldsymbol{\omega}_{k(n)} \tag{6}$$

The matrix $\mathbf{B}^T_{1(n,n)}$ is square and usually invertible. Hence it is possible, to express the previous system of equations in a staggered from by inverting the $\mathbf{B}^T_{1(n,n)}$. In fact, solving eq. (6) with respect to $\mathbf{n}_1$ and denoting for conciseness the vector of the axial hyperstatic strains $\mathbf{n}_{2\,(N-n)}$ as $\mathbf{y}_{(N-n)}$ is obtained:

$$\mathbf{n}_{(N)} = \begin{bmatrix} \mathbf{n}_{1(n)} \\ \mathbf{n}_{2(N-n)} \end{bmatrix} = \begin{bmatrix} \left(\mathbf{B}^T_{1(n,n)}\right)^{-1} \sum_{k=1}^{3} f_k \boldsymbol{\omega}_{k(n)} - \left(\mathbf{B}^T_{1(n,n)}\right)^{-1} \mathbf{B}^T_{2(n,N-n)} \mathbf{y}_{(N-n)} \\ \mathbf{y}_{(N-n)} \end{bmatrix} \tag{7}$$

Setting further:

$$\mathbf{H}_{(N,N-n)} = \begin{bmatrix} -\left(\mathbf{B}^T_{1(n,n)}\right)^{-1} \mathbf{B}^T_{2(n,N-n)} \\ \mathbf{I}_{(N-n,N-n)} \end{bmatrix}; \quad \mathbf{n}_{0(N)} = \begin{bmatrix} \left(\mathbf{B}^T_{1(n,n)}\right)^{-1} \sum_{k=1}^{3} f_k \boldsymbol{\omega}_{k(n)} \\ \mathbf{0}_{(N-n)} \end{bmatrix} \tag{8}$$

in which $\mathbf{I}_{(N-n,N-n)}$ is the identity matrix and $\mathbf{0}_{(N-n)}$ is the null vector, The vector n(n) can me expressed as:

$$\mathbf{n}_{(N)} = \begin{bmatrix} \mathbf{n}_{1(n)} \\ \mathbf{n}_{2(N-n)} \end{bmatrix} = \begin{bmatrix} \left(\mathbf{B}_{1(n,n)}^T\right)^{-1} \sum_{k=1}^{3} f_k \boldsymbol{\omega}_{k(n)} \\ \cdots \\ \mathbf{0}_{(N-n)} \end{bmatrix} + \begin{bmatrix} -\left(\mathbf{B}_{1(n,n)}^T\right)^{-1} \mathbf{B}_{2(n,N-n)}^T \\ \cdots \\ \mathbf{I}_{(N-n,N-n)} \end{bmatrix} \mathbf{y}_{(N-n)} = \quad (9)$$

$$= \mathbf{n}_{0(N)} + \mathbf{H}_{(N,N-n)} \cdot \mathbf{y}_{(N-n)};$$

In this way the axial strain vector is expressed as the sum of a known vector $\mathbf{n}_{0(N)}$, in balance with the nodal and an additional distribution of axial strains, depending the forces hyperstatic unknowns. The known axial strain $\mathbf{n}_{0(N)}$ may be conveniently regarded as the sum of the axial strains in the beans associated with for each distortive parameter:

$$\mathbf{n}_{0(N)} = \begin{bmatrix} \left(\mathbf{B}_{1(n,n)}^T\right)^{-1} \sum_{k=1}^{3} f_k \boldsymbol{\omega}_{k(n)} \\ \mathbf{0}_{(N-n)} \end{bmatrix} =$$

$$= \begin{bmatrix} \left(\mathbf{B}_{1(n,n)}^T\right)^{-1} f_1 \boldsymbol{\omega}_{1(n)} \\ \mathbf{0}_{(N-n)} \end{bmatrix} + \begin{bmatrix} \left(\mathbf{B}_{1(n,n)}^T\right)^{-1} f_2 \boldsymbol{\omega}_{2(n)} \\ \mathbf{0}_{(N-n)} \end{bmatrix} + \begin{bmatrix} \left(\mathbf{B}_{1(n,n)}^T\right)^{-1} f_3 \boldsymbol{\omega}_{3(n)} \\ \mathbf{0}_{(N-n)} \end{bmatrix} \quad (10)$$

Thus, setting:

$$\mathbf{f}_{k(N)} = \begin{bmatrix} \left(\mathbf{B}_{1(n,n)}^T\right)^{-1} \boldsymbol{\omega}_{k(n)} \\ \mathbf{0}_{(N-n)} \end{bmatrix} \quad (11)$$

one has

$$\mathbf{n}_{0(N)} = \sum_{k=1}^{3} f_k \mathbf{f}_{k(N)} \quad (12)$$

and eq. (12) eq. (9) can be written as:

$$\mathbf{n}_{(N)} = \sum_{k=1}^{3} f_k \mathbf{f}_{k(n)} + \mathbf{H}_{(N,N-n)} \mathbf{y}_{(N-n)} \quad (13)$$

In conclusion the axial strain $n_i$ in the *i-th* bean can be expressed as:

$$n_i = \sum_{k=1}^{3} f_k f_{ki} + \sum_{j=1}^{N-n} h_{ij} y_j \quad (14)$$

In order to enforce collimation between the observed energy and the computed one we need to introduce. The elastic elongation in the i-th beam is introduced and defined as:

$$\Delta l_i = \sum_{j=1}^{n} b_{ij} u_j \tag{15}$$

Where $b_{ij}$ denotes the entries of the compatibility matrix **B**.

$$\mathbf{\Delta l}_{(N)} = \mathbf{B}_{(N,n)} \mathbf{u}_{(n)} \tag{16}$$

## 5  The "object" energy distribution

Introducing the matrix of connection matrix $\mathbf{C}_{(\bar{n},N)}$ whose generic entry is defined by:

$$c_{ij} = \begin{cases} 1 & \Leftrightarrow \text{if the } j \text{ edge of the network is connected with the } i \text{ node} \\ 0 & \Leftrightarrow \text{other wise} \end{cases} \tag{17}$$

Assuming that the elastic energy of each beam equally splits between the two nodes to which it is connected, the energy at each node is given by the well known *Clapeyron* formula:

$$x_j = \frac{1}{4} N_j \Delta l_j \tag{18}$$

The energetic collimation between the observed energy $\varphi_{(\bar{n})}$ and the computed one $\mathbf{x}_{(N)}$ is expressed as:

$$\mathbf{C}_{(\bar{n},N)} \mathbf{x}_{(N)} = \varphi_{(\bar{n})} \tag{19}$$

In particular, the energetic comparison at the *i-th* node reads:

$$\varphi_i = \sum_{j=1}^{N} c_{ij} x_j \tag{20}$$

Eq. (19) of energetic collimation is a system of "$\bar{n}$" equations in the "$N$" unknowns the vector $\mathbf{x}_{(N)}$, in general the number of the unknowns is greater than the number of the equations making the problem hyperstatic. In order to compute the hyperstatic unknowns it is convenient to partition the connection matrix as follows:

$$\mathbf{C}_{(\bar{n},N)} = \left[ \mathbf{C}_{1(\bar{n},\bar{n})} \vdots \mathbf{C}_{2(\bar{n},N-\bar{n})} \right] \tag{21}$$

where $\mathbf{C}_{1\,(\bar{n},\bar{n})}$ is a non-singular square matrix and $\mathbf{C}_{2\,(\bar{n},N-\bar{n})}$ is the complementary rectangular matrix $\mathbf{C}_{1\,(\bar{n},\bar{n})}$. The same operation may be performed on the calculated energy vector by splitting **x** in the form:

$$\mathbf{x}_{(N)} = \begin{bmatrix} \mathbf{x}_{1(\bar{n})} \\ \mathbf{x}_{2(N-\bar{n})} \end{bmatrix} \tag{22}$$

Thus, the energetic collimation equation can be reformulated (19) by writing:

$$\begin{bmatrix} \mathbf{C}_{1(\bar{n},\overline{n0})} & \mathbf{C}_{2(\bar{n},N-\bar{n})} \end{bmatrix} \begin{bmatrix} \mathbf{x}_{1(\bar{n})} \\ \mathbf{x}_{2(N-\bar{n})} \end{bmatrix} = \varphi_{(\bar{n})} \tag{23}$$

The solution of the previous systems of equation is searched by pivoting on $\mathbf{x}_{1(\bar{n})}$ by introducing the vector **t** of the free energetic collimation variables:

$$\mathbf{x}_{(N)} = \begin{bmatrix} \mathbf{x}_{1(\bar{n})} \\ \mathbf{x}_{2(N-\bar{n})} \end{bmatrix} = \begin{bmatrix} \left(\mathbf{C}_{1(\bar{n},\bar{n})}\right)^{-1} \varphi_{(\bar{n})} - \left(\mathbf{C}_{1(\bar{n},\bar{n})}\right)^{-1} \mathbf{C}_{2(\bar{n},N-\bar{n})} \mathbf{t}_{(N-\bar{n})} \\ \mathbf{t}_{(N-\bar{n})} \end{bmatrix} \tag{24}$$

Thus the first "$\bar{n}$" energy unknowns can be expressed as function of the remaining "$N-\bar{n}$" unknowns. In particular, setting:

$$\mathbf{L}_{(N,N-\bar{n})} = \begin{bmatrix} -\left(\mathbf{C}_{1(\bar{n},\bar{n})}\right)^{-1} \mathbf{C}_{2(\bar{n},N-\bar{n})} \\ \mathbf{I}_{(N-\bar{n},N-\bar{n})} \end{bmatrix} \quad \mathbf{x}_{0(N)} = \begin{bmatrix} \left(\mathbf{C}_{1(\bar{n},\bar{n})}\right)^{-1} \varphi_{(\bar{n})} \\ \mathbf{0}_{(N-\bar{n})} \end{bmatrix} \tag{25}$$

in which $\mathbf{L}_{(N,N-\bar{n})}$ is the matrix of the free variables of energetic collimation **t**, $\mathbf{I}_{(N-\bar{n},N-\bar{n})}$ is the identity matrix and **0** is the null vector, the vector $\mathbf{X}_{(N)}$ can be written like:

$$\mathbf{x}_{(N)} = \begin{bmatrix} \mathbf{x}_{1(\bar{n})} \\ \mathbf{x}_{2(N-\bar{n})} \end{bmatrix} = \begin{bmatrix} \left(\mathbf{C}_{1(\bar{n},\bar{n})}\right)^{-1} \varphi_{(\bar{n})} \\ \mathbf{0}_{(N-\bar{n})} \end{bmatrix} + \begin{bmatrix} -\left(\mathbf{C}_{1(\bar{n},\bar{n})}\right)^{-1} \mathbf{C}_{2(\bar{n},N-\bar{n})} \\ \mathbf{I}_{(N-\bar{n},N-\bar{n})} \end{bmatrix} \mathbf{t}_{(N-\bar{n})} = \tag{26}$$

$$= \mathbf{x}_{0(N)} + \mathbf{L}_{(N,N-\bar{n})} \mathbf{t}_{(N-\bar{n})}$$

This operation has the purpose of isolating the hyperstatic variables, in the sense that the vector of the beam energies is expressed as the sum of a known vector $\mathbf{x}_{0(N)}$, obtained from the observed energies, and from a further energetic distribution dependent on the free variable unknowns of energetic collimation **t**.

## 6 Objective function

The unknowns of the problem are actually evaluated by searching the minimum of the *average square error* E between the calculated energies and those observed in each beam; on account of the equilibrium equation (2) and compatibility one (15) the average square can be expressed:

$$E(\mathbf{y},\mathbf{u},\mathbf{t},\mathbf{f}) = \sum_{i=1}^{N}\left[0.25 N_i \Delta l_i - x_i\right]^2 = \left[\mathbf{x}_{(N)} - \left(\mathbf{x}_{o(N)} + \mathbf{L}_{(N,N-n)}\mathbf{t}_{(N-n)}\right)\right]^2 \quad (27)$$

Whose minimum reached is under the condition ensuring the positivity of the rigidity of the *i-th* beam:

$$k_i = \frac{N_i}{\Delta l_i} > 0 \Rightarrow N_i \Delta l_i > 0 \quad (28)$$

As a matter of fact, the solution is obtained by solving a *structural optimization problem* (Non-Linear bounded Programming, **NLP**) for a given geometry.

In conclusions, the mathematical problem is to solve amounts to evaluate the minimum of non-linear function of several variables subject to nonlinear constraints. Among several procedures available in the literature, the *Penalization method* is selected. Since it amounts to search the constrained minimum by solving a unconstrained optimization problem based on suitably modified objective function that incorporates the constraints as a penalization term. For the case at hand thus term reads so that the unconstrained minimum of the functional:

$$P(\mathbf{y},\mathbf{u},\mathbf{f}) = \sum_{i=1}^{N}\left(N_i \Delta l_i\right)^2 \quad (29)$$

Therefore, researching the free minimum of the function:

$$G(\mathbf{y},\mathbf{u},\mathbf{t},\mathbf{f}) = E(\mathbf{y},\mathbf{u},\mathbf{t},\mathbf{f}) + rP(\mathbf{y},\mathbf{u},\mathbf{f}) \quad (30)$$

retrieves the constrained minimum of the functional $E$ when $r \to +\infty$

## 7 Formulization of the inverse problem

Having indentified the elastic proprieties of the *equivalent reticular structure* (mesh), it is possible to run the inverse procedure.

Once derived the distortional parameters **f** at the epicentre, an elastic analysis of the *mesh* is done, on the basis of the values of the calculated *optimal stiffness*:

$$\left[\underbrace{\mathbf{B}_{(n,N)}^{T}\mathbf{D}_{(N,N)}\mathbf{B}_{(N,n)}}\mathbf{u}_{(n)} = \mathbf{f}_{(n)} \Leftrightarrow \mathbf{K}_{(N,n)}\mathbf{u}_{(n)} = \mathbf{f}_{(n)}\right] \quad (31)$$

where $\mathbf{f}_{(n)}$ is the vector of the active loads (known):

$$\mathbf{f}_{(n)} = \sum_{k=1}^{3} \mathbf{f}_k \boldsymbol{\omega}_{k(n)} \tag{32}$$

$\mathbf{u}_{(n)}$ is the vector containing the unknown nodal displacements, with which the examined area has been discretized, matrix $\mathbf{D}_{(N,N)}$ contains the constitutive low of the material; $\mathbf{K}_{(N,n)}$ is the stiffness matrix and $\mathbf{B}_{(N,n)}$ is the compatibility matrix eq. (16). The solution to the eq. (31) is achieved by inverting the matrix $\mathbf{K}_{(N,n)}$ arriving at the following relation:

$$\mathbf{u}_{(n)} = \mathbf{K}_{(N,n)}^{-1} \mathbf{f}_{(n)} \tag{33}$$

By which, the elastic energy stored is obtained in each node of the *equivalent reticular structure* through:

$$\varphi_{C(n)} = \frac{1}{4}\left(\mathbf{C}_{(n,N)} diag\left(\mathbf{n}_{(N)} \Delta \mathbf{l}_{(N)}^T\right)\right) = \frac{1}{4}\left(\mathbf{C}_{(n,N)} \mathbf{K}_{(N,N)} \left(\Delta \mathbf{l}_{(N)}\right)^2\right) \tag{34}$$

In which:

$$\Delta \mathbf{l}_{(N)} = \mathbf{B}_{(N,n)} \cdot \mathbf{u}_{(n)} \tag{35}$$

Assuming that, the vector of the nodal elastic energy is known, nodal intensities can be evaluated through the inverse of the eq. (1):

$$\mathbf{I}_{C(n)} = 5 + \log_{10} \varphi_{o(n)} \tag{36}$$

## 8  Optimization

The problem is developed assuming the following *Mono Objective Optimization* of the four events considered:

$$\min G_e(\mathbf{y},\mathbf{u},\mathbf{t},\mathbf{f}) = \min\left[E(\mathbf{y},\mathbf{u},\mathbf{t},\mathbf{f}) + rP(\mathbf{y},\mathbf{u},\mathbf{f})\right]_e =$$

$$= \min\left[\sum_{i=1}^{N}\left(\frac{1}{4}\mathbf{n}_i \Delta l_i - x_i\right)^2 + r\sum_{i=1}^{N}\left(\mathbf{n}_i \Delta l_i\right)^2\right]_e \tag{37}$$

With $e=event=1,..., 4$; by which the *distortive parameters of* the four epicentre cells are assumed known, referred to the events considered.
The optimization of the eq. (37) is carried out by the *Descent Methods*, iterative methods, which, as known, employ information on the inclination $\nabla f(\mathbf{x})$ of the objective function $f(\mathbf{x})$ to determine a direction of research, $\mathbf{d}_k$. Such methods have a major computation weight, because the necessary operations during each iteration are more complex than those carried out by methods that do not utilize derivations. Nevertheless, the latter are unable to guarantee the same convergence characteristics, and normally a greater number of iterations are necessary to obtain a solution (the number of iterations is usually very high and sometimes, in order to reach the optimum, said number tends towards the infinite).

What distinguishes one method of descent from another, are the criteria used to choose the direction of descent, $\mathbf{d}_k$ and the length of the step, $\alpha_k$.

The direction of research may be determined by considering a suitable approximation of the objective function regarded as function of the vector direction of optimizing, $d_k$. In the numeric calculation code, developed *ad- hoc*, the function approximation is of quadratic type with continuous Hessian, symmetric and defined positive.

By the Taylor formula, terminated at the 2nd order terms for sufficiently low values of the vector increment rule $d_k$:

$$\min_{\mathbf{x}} f(\mathbf{x}_k + \mathbf{d}_k) \approx \min_{\mathbf{x}} \left( f(\mathbf{x}_k) + \nabla f(\mathbf{x}_k)^T \mathbf{d}_k + \frac{1}{2} \mathbf{d}_k^T \nabla^2 f(\mathbf{x}_k) \mathbf{d}_k \right) \qquad (38)$$

The optimal solution is obtained when the partial derivatives are null:

$$\nabla f(\mathbf{d}_k) = \nabla f(\mathbf{x}_k)^T + \nabla^2 f(\mathbf{x}_k) \mathbf{d}_k = 0 \qquad (39)$$

The optimal point in solution may be written as:

$$\mathbf{d}_k = -\left(\nabla^2 f(\mathbf{x}_k)\right)^{-1} \nabla f(\mathbf{x}_k) \qquad (40)$$

The method consists in fixing, for the direction, precisely $\mathbf{d}_k$, and along this direction $\alpha_k$ is moved by one step. The problem regarding the determination of the step, $\alpha_k$, in direction of the descent, $\mathbf{d}_k$, takes the name of line *search* (as it happens along a "line", or the direction of descent). The problem is to find a new iteration of the form

$$\mathbf{x}_{k+1} = \mathbf{x}_k + \alpha_k^* \mathbf{d}_k \qquad (41)$$

Where $\mathbf{x}_k$ denotes the current iteration, $\mathbf{d}_k$ is the direction of descent, and $\alpha_k^*$ is a parameter of the step's length (scaling), which is the distance from the minimum.

The minimum along the line formed by the descent direction is generally estimated by using a research procedure, or by a polynomial method which involves extrapolation or interpolation. In the implemented calculation code, the minimum has been estimated by using the polynomial cubic interpolation method. The basic idea of interpolation is that of utilizing an approximated third-degree polynomial representation of the step in the interval of the values of interest, which are those included between 0 and the current step.

It is known that the *Quasi-Newton* method avoids the numeric calculation of the Hessian matrix (*Newton Method*), by utilising the observed behaviour of the objective function $f(\mathbf{x})$ and of the gradient $\nabla f(\mathbf{x})$ in order to develop information on the curvature to make an approximation of the Hessian matrix, using appropriate updating techniques.

In literature, a great number of updating-inversion methods of the Hessian matrix have been developed. Among these, two methods have been chosen: BFGS and DFP.

In the BFGS method (Broyden 1980, Fletcher 1970, Godfarb 1970 and Shanno 1970), the matrix represent by the equation (42):

$$\left[\nabla^2 f(\underline{\mathbf{x}}_{k+1})\right]^{-1} = \mathbf{H}_{k+1}^{-1} = \mathbf{H}_k^{-1} + \frac{\mathbf{q}_k \mathbf{q}_k^T}{\mathbf{q}_k^T \mathbf{s}_k} - \frac{\left(\mathbf{H}_k^{-1}\right)^T \mathbf{s}_k^T \mathbf{s}_k \mathbf{H}_k^{-1}}{\mathbf{s}_k^T \mathbf{H}_k^{-1} \mathbf{s}_k}$$

where:
$$\mathbf{s}_k = \mathbf{x}_{k+1} - \mathbf{x}_k$$
$$\mathbf{q}_k = \nabla f(\mathbf{x}_{k+1}) - \nabla f(\mathbf{x}_k)$$

(42)

while the DFP method (Davidon 1959, Fletcher and Powell 1963) is similar with the previous one by replacing the qk by sk.

The research direction is determined in automatic by code - by a choice between the BFGS method and the DFP method. The Hessian matrix always remains positive, while the research direction, dk, is always in the direction of descent. The positivity of the Hessian is realized by assuring that it is initialized to be defined positive and that, thereafter, the following term will always be positive:

$$\mathbf{q}_k^T \mathbf{s}_k = \alpha_k^* (\nabla f(\mathbf{x}_{k+1})^T \mathbf{d}_k - \nabla f(\mathbf{x}_k)^T \mathbf{d}_k)$$

(43)

In the polynomial cubic method proposed at each iteration the gradient and the function are calculated, the updating of the Hessian is accomplished when the new point found:

$$\mathbf{x}_{k+1} = \mathbf{x}_k + \alpha_k^* \mathbf{d}_k$$

(44)

meets the condition:

$$f(\mathbf{x}_{k+1}) < f(\mathbf{x}_k)$$

(45)

That is, when the method is actually a method of descent. If such a condition doesn't occur, then the step $\alpha_k^*$ is reduced to form a new step $\alpha_{k+1}$.

## 9  Numeric results: mono-objective optimization

Shown below are the results obtained by the proposed modelling, focalized on the four events under exam and illustrated in Table 2.

Table 2 - *Four selected events registered in historical dates within the area under study*

| Year | Day | Month | Ma [s] | Io [MCS×10] | Imax [MCS×10] |
|------|-----|-------|--------|-------------|---------------|
| 1913 | 4   | October | 5.15 | 75 | 80 |
| 1930 | 23  | July    | 6.72 | 100 | 100 |
| 1962 | 21  | August  | 6.19 | 90 | 90 |
| 1980 | 23  | November| 6.89 | 100 | 100 |

- The 4 October 1913 earthquake

The map of the isosistes superimposed on the equivalent reticular structure (*mesh*) is represented in Figure 4a. The comparison between the values of the calculated intensities and the observed intensities in the various knots of the *mesh* is shown in Figure 4b. The nodal displacements obtained from the optimization is shown in Figure 4c.

We have compared the values of the *classes of calculated intensity* in each *cell* of the *mesh*, obtained from the average of the intensity values of the four bordering knots, with the observed macro-seismic field, Figure 4d, and with the observed isosistes, Figure 4e. In addition, Figure 4f shows the trend of the intensity residues.

- The 23 July 1930 earthquake

Figure 5a shows the Isosistes map, overlaid on the equivalent reticular structure (*mesh*); Figure 5b represents the comparison between the values of the calculated intensities and the observed intensities in the various knots of the *mesh*; in Figure 5c the nodal displacements obtained by optimization is shown. We have compared the values of the *classes of intensity* calculated within each *cell* of the *mesh*, obtained as an average of the values of the intensities of the four surrounding knots, with the observed macro-seismic field, in Figure 5d, and with the observed Isoseists, in Figure 5e. Figure 5f shows the intensity residue trends.

- The 21 August 1962 earthquake

We have represented in Figure 6a the Isosistes map, overlaid on the equivalent reticular structure (*mesh*); in Figure 6b the comparison between the calculated intensities and the observed intensities in the various mesh knots; in Figure 6c the nodal shifting obtained by optimization. We compared the values of the *intensity classes* calculated in each *meshcell*, obtained as the average of the intensity values of the four adjacent knots, with the observed macro-seismic field, Figure 6d, and with the observed Isoseistes, Figure 6e. Figure 6f shows the trend of the intensity residues.

- The 23 November 1980 earthquake

We represent in Figure 7a the Isosistes map overlaid on the equivalent reticular structure (mesh); in Figure 7b the comparison between the calculated intensities and the observed intensities in the various mesh knots; in Figure 7c the nodal shifting obtained by optimizing. We compared the values of the classes of intensities calculated in each mesh cell, obtained as the average of the intensity values of the four adjacent knots with the map of the observed Isosistes, Figure 7d, and with the observed macro-seismic field, Figure 7e. Figure 7f shows the trend of the intensity residues.

The research is mainly aimed at assessing seismic risk at territorial scale (Zuccaro and De Gregorio, 2013; Zuccaro et al., 2017), but it is also useful for the development of new construction technologies and materials in seismic engineering (Cavalieri et al., 2017; Pingue et al., 2015).

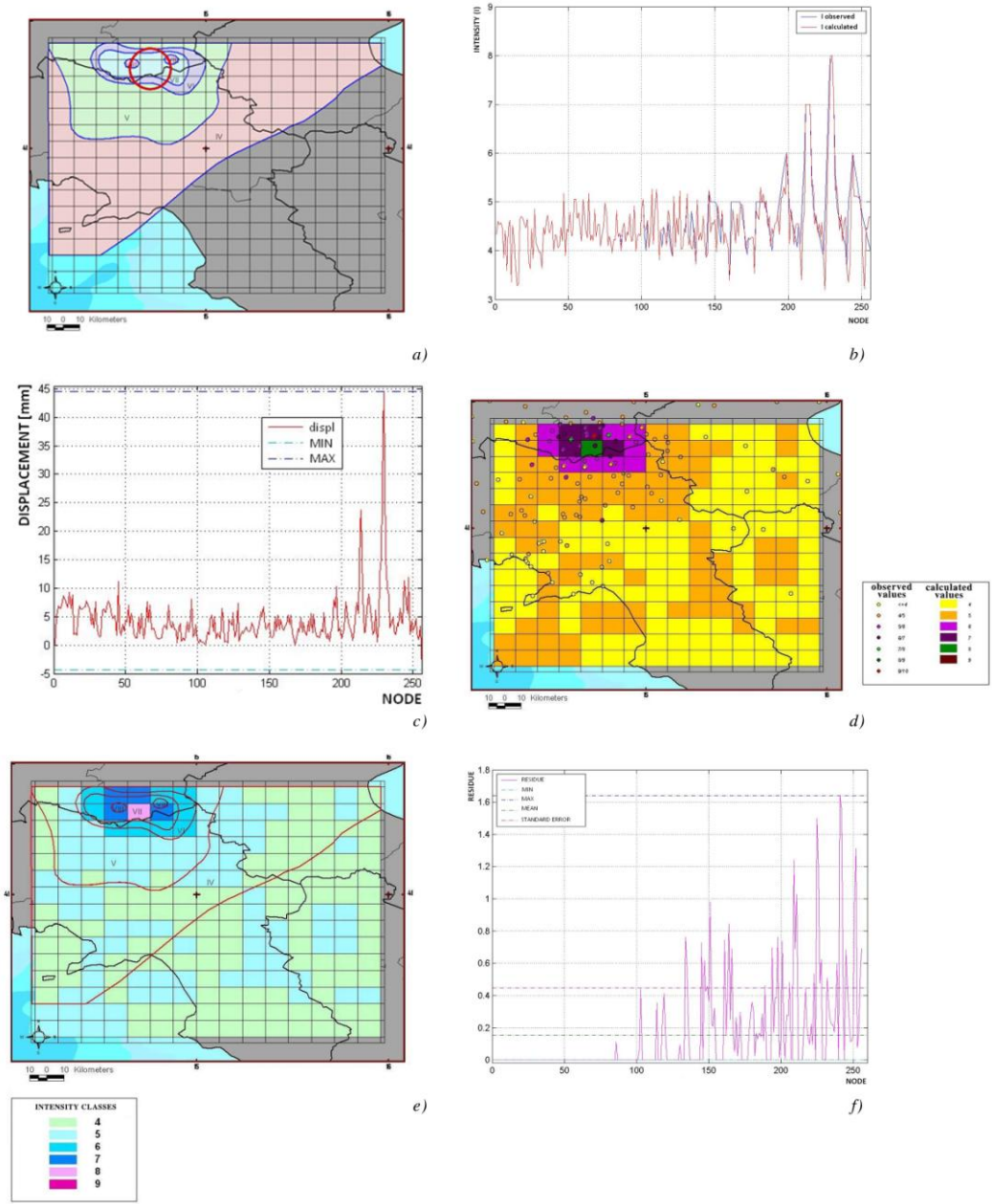

Figure 4: *a) Isosistes Map of the 4/10/1913 earthquake superimposed to on the mesh (the epicentre cell is circled in red), b) Comparison between observed and computed nodal intensities, c) Shifting of the mesh nodes, d) Macro-seismic field compared with the intensity classes generated by the model, e) Isosistes Map compared with the intensity classes generated by the model, f) Intensity residue trend.*

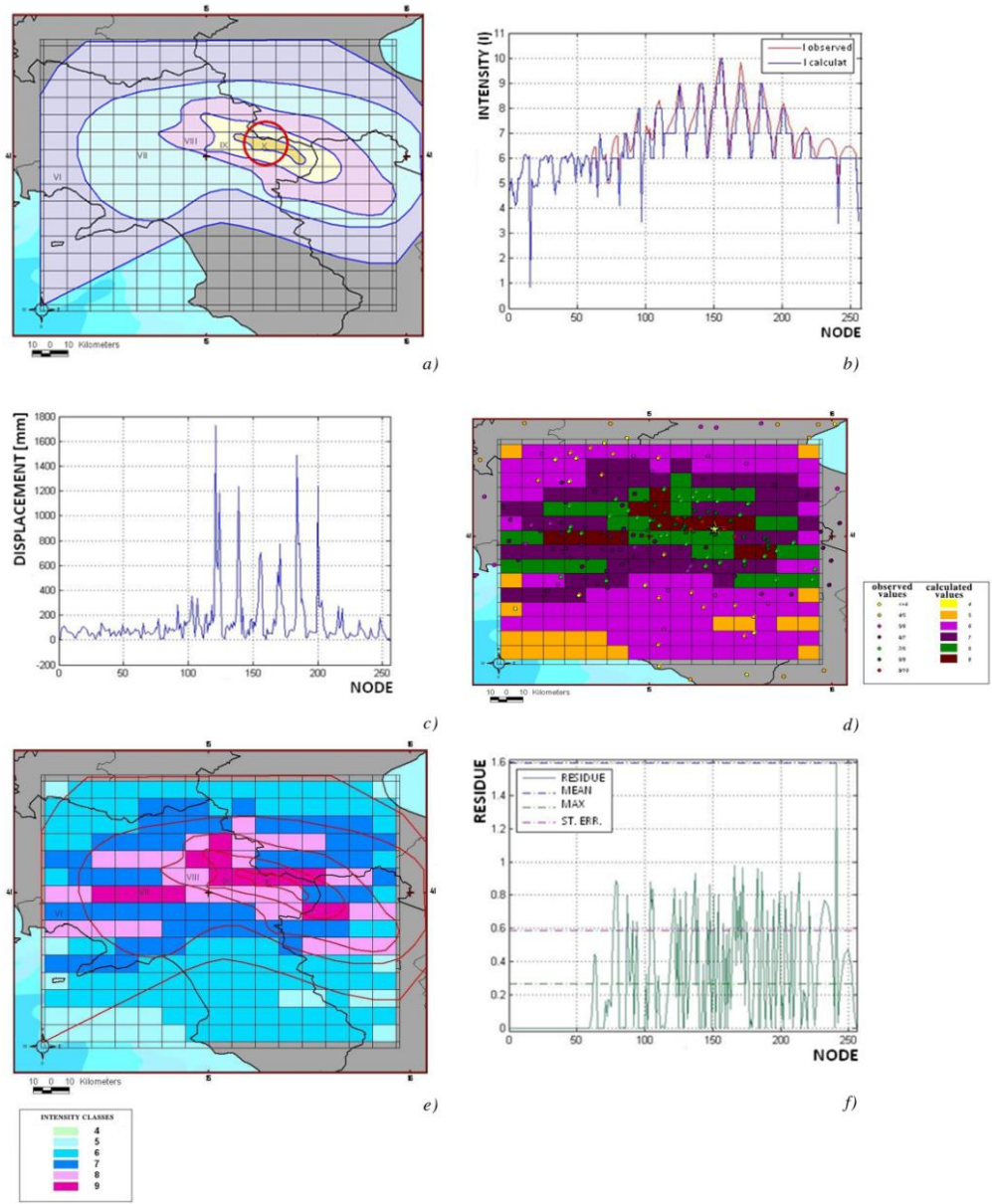

Figure 5: *a) Isosistes Map of the 23/07/1930 earthquake superimposed to on the mesh, b) Comparison between observed and computed nodal intensities, c) Shifting of the mesh nodes, d) Macro-seismic field compared with intensity classes generated by the model, e) Isosistes Map compared with the intensity classes generated by the model, f) Intensity residue trend.*

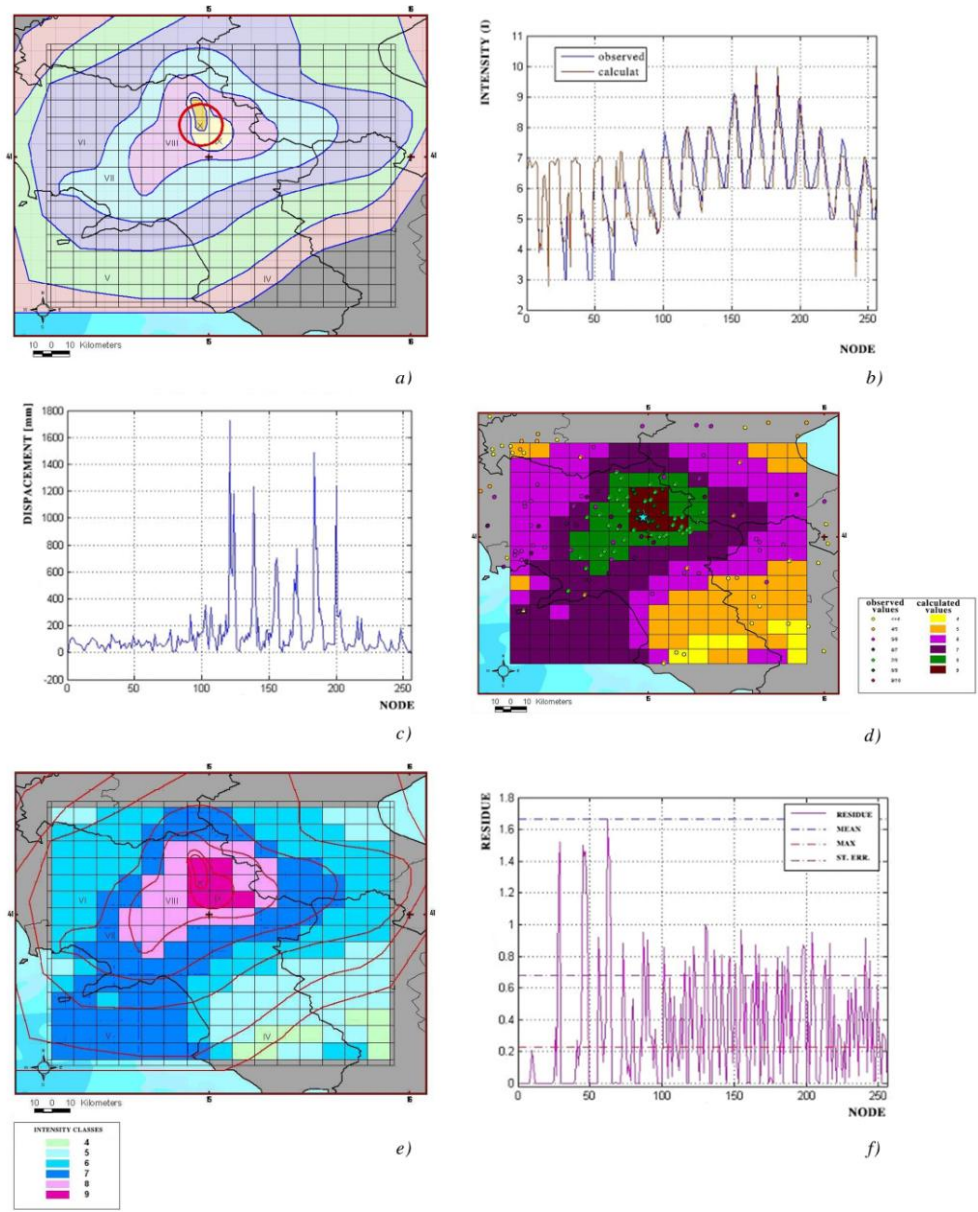

Figure 6: *a) Isosistes Map of the 21/08/1962 earthquake superimposed to on the mesh, b) Comparison between observed and computed nodal intensities, c) Shifting of the mesh nodes, d) Macro-seismic field compared with intensity classes generated by the model, e) Isosistes Map compared with the intensity classes generated by the model, f) Intensity residue trend.*

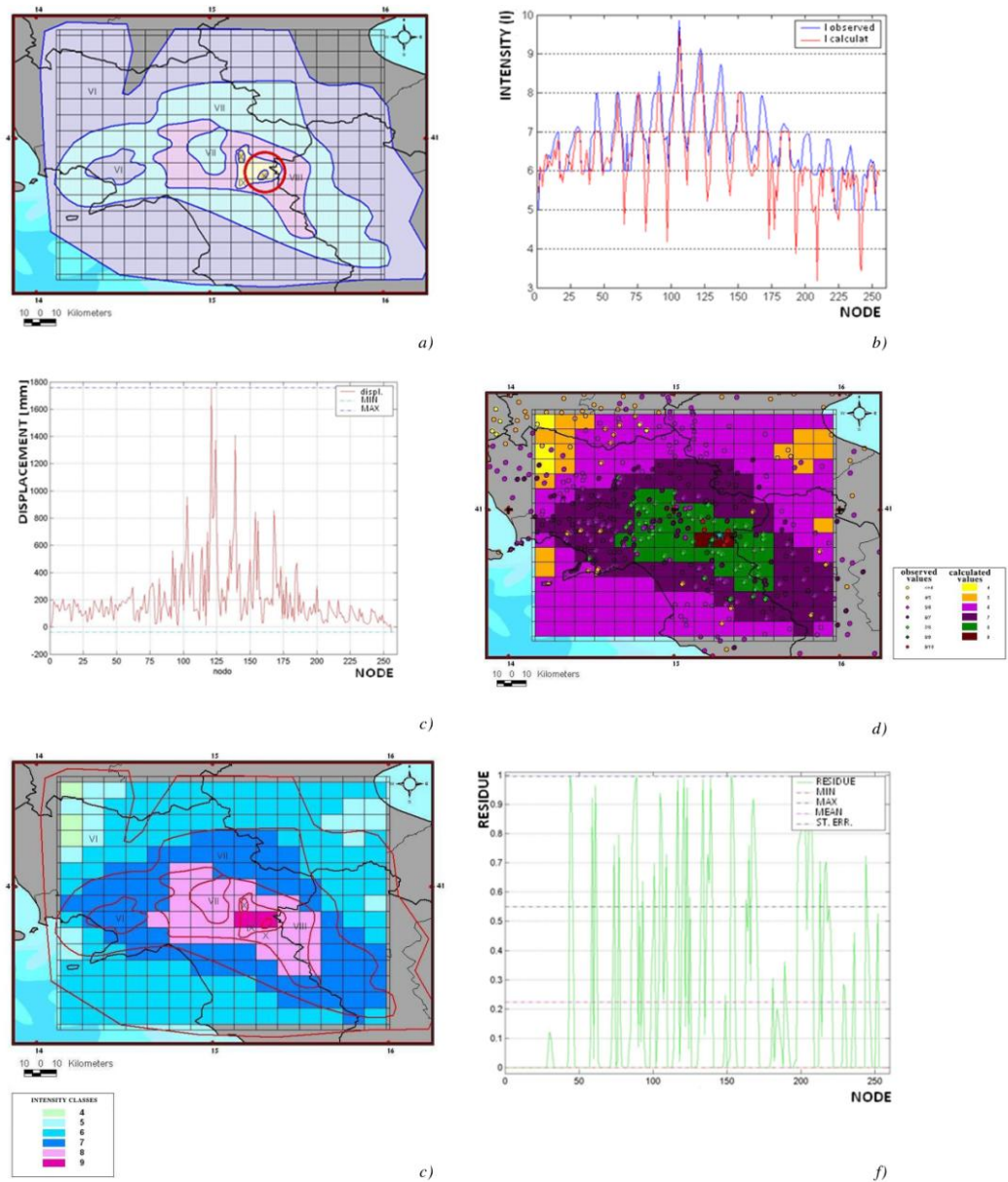

Figure 7: *a) Isosistes Map of the 23/11/1980 earthquake superimposed to on the mesh, b) Comparison between observed and computed nodal intensities, c) Shifting of the mesh nodes, d) Macro-seismic field compared with intensity classes generated by the model, e) Isosistes Map compared with the intensity classes generated by the model, f) Intensity residue trend.*

## Conclusions

The purpose of this work is to develop a novel methodology able to reproduce the propagation law of seismic energy, hypothesizing the "transmission" mechanisms that preside over the distribution of seismic effects on the territory, by means of a structural optimization process with a predetermined energy distribution. In order to achieve this goal, the present paper proposes a physical mechanical prototype of seismic energy transmission in the Campania territory.

The results obtained have shown promising results; the nodal values of the intensities tend to show a good accordance, the intensities observed less than 5-6 are those that differ most from the intensities generated by the model, due probably to the bigger uncertainties tied to the low intensities.

Through the comparison of the classes of intensity, calculated according to the map of observed Isosistes and the field of macro-seismic observations, we note the close approximation of the intensities generated by the model and therefore, a satisfactory grade of approximation of the simulation technique proposed in each single event.

We also note, from the trend of the intensity residuals, that there is a close grade of approximation between the observed values and the calculated ones; in fact, there isn't a tendency to growth of the residuals and they are all considerably contained.

The results of the proposed model have particular interest, both as an aid in the defining of local laws of the attenuation of seismic energy let out at the epicentre and as an eventual analysis of the evaluation of local geological effect at macro-scale. The latter is of great actuality concerning the topics of seismic reclassification on a regional scale.

For the future developments of the present research, the authors propose a procedure based on *multi-objective* analysis, having the capacity of averaging, for the four selected earthquakes, the characteristics of stiffness of the soil according to a star of directions.

# ON THE SIMULATION OF THE SEISMIC ENERGY TRANSMISSION MECHANISMS

*Giulio Zuccaro[1,2], Daniela De Gregorio[2], Magdalini Titirla[3], Mariano Modano[1], Luciano Rosati[1]*

[1]Department of Structures for Engineering and Architecture, University of Naples Federico II, Napoli, Italy

[2]LUPT-PLINIVS Study Centre, University of Naples Federico II, Napoli, Italy

[3]Laboratory of Composite Materials for Construction LMC2, University Claude Bernard Lyon 1, Lyon, France

SOMMARIO: *Oggi, i metodi in grado di valutare la propagazione dell'energia sismica sul territorio assumono una notevole rilevanza. La propagazione dell'energia sismica è fortemente correlata alla complessità della sorgente ed è influenzata dall'attenuazione e dagli effetti di dispersione lungo il percorso. Pertanto, l'effetto del terremoto sul territorio è il risultato di una complessa interazione tra il segnale emesso dalla sorgente e gli effetti di propagazione.*

*Lo scopo di questa ricerca è quello di sviluppare una metodologia in grado di riprodurre la legge di propagazione dell'energia sismica, ipotizzando i meccanismi di "trasmissione" alla base della distribuzione degli effetti sismici sul territorio, attraverso un processo di ottimizzazione strutturale rispetto ad una distribuzione di energia predeterminata.*

*In breve, l'approccio, basato su un modello fisico di tipo deterministico, valuta una correzione oggettiva delle distribuzioni rilevate di intensità sismica sul territorio, imponendo la compatibilità dei dati osservati con il modello fisico-meccanico. Quest'ultimo si basa su due ipotesi: (1) il terremoto all'epicentro è simulato mediante un sistema di distorsioni caratterizzato da tre parametri; (2) l'intensità è considerata coincidente con la densità dell'energia elastica. La distribuzione ottimale della rigidezza dei link si ottiene riducendo la differenza tra i valori di distribuzione dell'intensità calcolati sulla rete e quelli osservati durante quattro eventi regionali occorsi in passato e riguardanti la regione Campania (Italia).*

KEYWORDS: *trasmissione di energia, metodo degli elementi finiti, procedura di ottimizzazione, rischio sismico, valutazione del rischio sismico.*

___________________
Corresponding author: Giulio Zuccaro, Department of Structures for Engineering and Architecture, University of Naples Federico II, Napoli, Italy.
Email: zuccaro@unina.it